\documentclass[preprint,preprintnumbers, prd, floatfix, superscriptaddress,nofootinbib] {revtex4-1}
\usepackage{epsfig}
\usepackage{subfigure}
\usepackage{dcolumn}
\usepackage{bm}
\usepackage[usenames ,dvipsnames]{xcolor}
\usepackage{slashed}
\usepackage{graphicx,color}

\begin{document}
\title{Non-leptonic two-body weak decays of $\Lambda_c(2286)$}

\author{C.Q. Geng}

\affiliation{School of Physics and Information Engineering, Shanxi Normal University, Linfen 041004, China}
\affiliation{Department of Physics, National Tsing Hua University, Hsinchu, Taiwan 300}

\author{Y.K. Hsiao}
\affiliation{School of Physics and Information Engineering, Shanxi Normal University, Linfen 041004, China}
\affiliation{Department of Physics, National Tsing Hua University, Hsinchu, Taiwan 300}

\author{Yu-Heng Lin}
\affiliation{Department of Physics, National Tsing Hua University, Hsinchu, Taiwan 300}

\author{Liang-Liang Liu}
\affiliation{School of Physics and Information Engineering, Shanxi Normal University, Linfen 041004, China}

\date{\today}

\begin{abstract}
We study 
 the non-leptonic two-body weak decays of $\Lambda_c^+(2286)\to {\bf B}_n M$
with ${\bf B}_n$ ($M$) representing as the baryon (meson) states. 
Based on the $SU(3)$ flavor symmetry, we can describe
most of the data reexamined by the BESIII Collaboration with higher precisions. 
However, our result of ${\cal B}(\Lambda_c^+ \to p\pi^0)=(5.6\pm 1.5)\times 10^{-4}$ is larger than the current experimental limit of $3\times10^{-4}$ (90\% C.L.) by BESIII. In addition, we find that ${\cal B}(\Lambda_c^+ \to \Sigma^+ K^0)=(8.0\pm 1.6)\times 10^{-4}$, ${\cal B}(\Lambda_c^+ \to \Sigma^+ \eta^\prime)=(1.0^{+1.6}_{-0.8})\times 10^{-2}$, and  ${\cal B}(\Lambda_c^+ \to p \eta^\prime)=(12.2^{+14.3}_{-\,\,\,8.7})\times 10^{-4}$, which are accessible to the BESIII experiments.
\end{abstract}

\maketitle
\section{introduction}
Recently, the BESIII Collaboration has reanalyzed 
the two-body weak decays of $\Lambda_c^+(2286)$
with the final states to be the combinations of
baryon (${\bf B}_n$) and  pseudoscalar meson ($M$) particles,
where $\Lambda_c^+\equiv \Lambda_c^+(2286)$ along with
 $\Xi_c^{+,0}(2470)$
belongs to the lowest-lying
anti-triplet charmed baryon (${\bf B}_c$) state.
In particular,
the decay branching ratios of 
$\Lambda_c^+ \to p \bar K^0, \Lambda \pi^+,\Sigma^+ \pi^0$ and 
$\Sigma^0\pi^+$ have been measured 
at the level of $10^{-2}$ with high precisions~\cite{Ablikim:2015flg}.
In addition, the Cabibbo-suppressed $\Lambda_c^+\to p\eta$ decay 
has been observed for the first time~\cite{Ablikim:2017ors}. 
%
According to the measurements of
the two-body $\Lambda_c^+\to{\bf B}_n M$ decays since 2016~\cite{Ablikim:2015flg},
there have been 4 measured branching fractions listed in PDG~\cite{pdg}, given as
\begin{eqnarray}\label{data1}
{\cal B} (\Lambda_c^+ \to p \bar K^0)
&=&(3.16\pm 0.16)\%\,,\nonumber\\
{\cal B} (\Lambda_c^+ \to \Lambda \pi^+)
&=&(1.30\pm 0.07)\%\,,\nonumber\\
{\cal B} (\Lambda_c^+ \to \Sigma^+ \pi^0)
&=&(1.24\pm 0.10)\%\,,\nonumber\\
{\cal B} (\Lambda_c^+ \to \Sigma^0 \pi^+)
&=&(1.29\pm 0.07)\%\,,
\end{eqnarray}
together with the new data~\cite{Ablikim:2017ors}, given by
\begin{eqnarray}\label{data1b}
{\cal B} (\Lambda_c^+ \to p \eta)
&=&(1.24\pm0.28\pm 0.10)\times10^{-3}\,,
\nonumber\\
{\cal B} (\Lambda_c^+ \to p \pi^0)
&<&3\times10^{-4}\;(\text{90\% C.L.})\,.
\end{eqnarray}
Note that 
the limit of ${\cal B} (\Lambda_c^+ \to p \pi^0)$ in Eq.~(\ref{data1b})
comes from the
original data
of ${\cal B}(\Lambda_c^+\to p\pi^0)=(7.95\pm 13.61)\times 10^{-5}$~\cite{ppi0} by BESIII,
while
the $\Lambda_c^+\to \Sigma^+ K^0,p\eta'$ and $\Sigma^+\eta'$ decays,
along with the neutron modes, have not been seen yet.
It is interesting to see if these current data can be understood.

Theoretically, the factorization approach is demonstrated to well explain 
the $B$ and $b$-baryon decays~\cite{ali,Geng:2006jt,Hsiao:2014mua},
such that it is also applied to 
the two-body $\Lambda_c^+\to {\bf B}_n M$ decays~\cite{Bjorken:1988ya},
of which the amplitudes are derived as the combination of 
the two computable matrix elements for
the $\Lambda_c^+\to{\bf B}_n$ transition and the meson ($M$) production.
However, 
the factorization approach does not work for most of the two-body
$\Lambda_c^+\to {\bf B}_n M$ ones. For example, 
the decays of 
$\Lambda_c^+\to \Sigma^+\pi^0$ and $\Xi^0 K^+$ are forbidden
in the factorization approach~\cite{Lu:2016ogy}, 
but their branching ratios turn out to be measured.
As a result, several theoretical attempts to improve the factorization
by taking into account the nonfactorizable effects
have been made~\cite{Cheng:1991sn,Cheng:1993gf,Zenczykowski:1993hw,Uppal:1994pt,Fayyazuddin:1996iy}.
In contrast with the QCD-based models,
the $SU(3)$ symmetry approach is independent of  the detailed dynamics,
which has been widely used in the $B$ meson~\cite{He:2000ys,Fu:2003fy,Hsiao:2015iiu}, 
$b$-baryon~\cite{He:2015fwa,He:2015fsa} and
$\Lambda_c^+$ ($\Xi_c$)~\cite{Lu:2016ogy,
Savage:1989qr,Savage:1991wu,Verma:1995dk,Sharma:1996sc} decays.
With this advantage,
the two-body $\Lambda_c^+\to {\bf B}_n M$ decays
can be related by the $SU(3)$ parameters, 
which receive possible non-perturbative and 
non-factorizable contributions~\cite{Sharma:1998rd,Lu:2016ogy,
Savage:1989qr,Savage:1991wu,Verma:1995dk,Sharma:1996sc},
despite of the unknown sources.
%
%
The minimum $\chi^2$ fit with the p-value estimation~\cite{pdg}
can statistically test if the $SU(3)$ flavor symmetry agrees with the data.
Being determined from the fitting also, 
the $SU(3)$ parameters are taken to
predict the not-yet-measured modes for the future experimental tests.
However, the global fit was once unachievable without the sufficient data and 
the use of the symmetry for $\Lambda_c^+\to {\bf B}_n M$.
Clearly,
the reexamination with the global fit to
match the currently more accurate data is  needed.
Note that, to study the $\Lambda_c^+\to{\bf B}_n\eta^{(\prime)}$ decays,
the singlet state of $\eta_1$ should be included~\cite{Fu:2003fy,Hsiao:2015iiu}.
%
In this report, we will extract the $SU(3)$ parameters in the global fit,
and predict the branching fractions to be compared with the future BESIII experimental measurements.

\section{Formalism}
\begin{figure}[t!]
\centering
\includegraphics[width=1.5in]{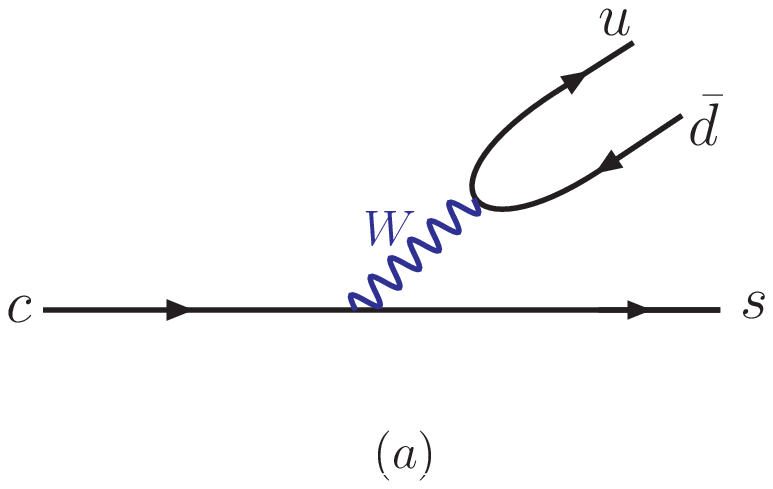}
\includegraphics[width=1.5in]{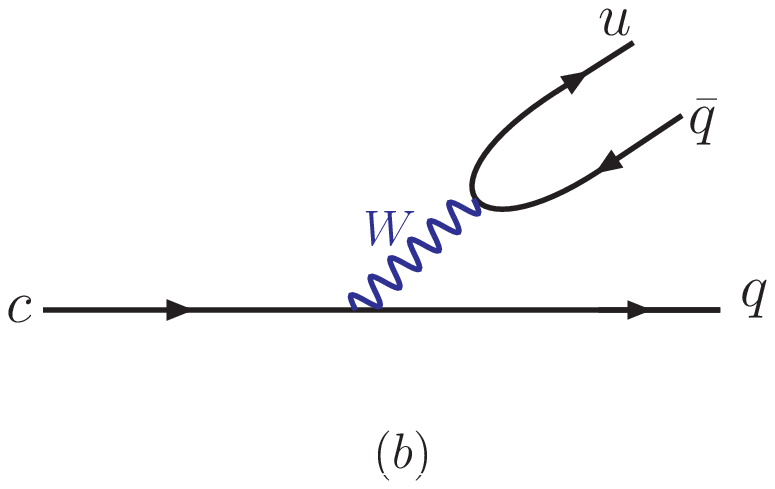}
\includegraphics[width=1.5in]{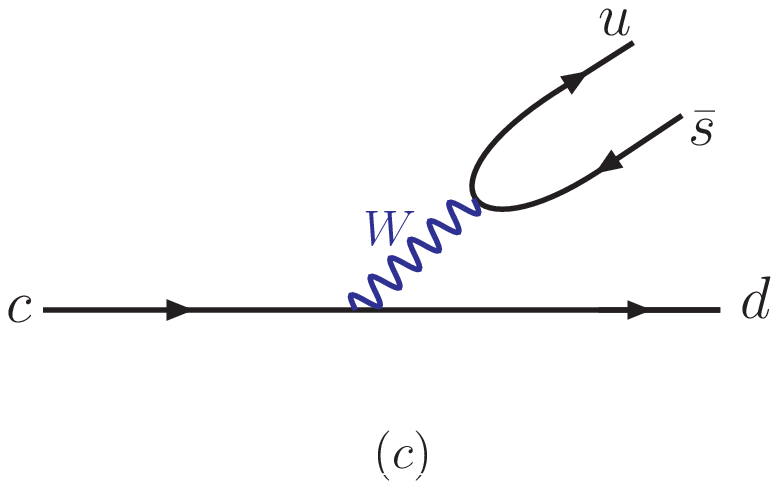}
\includegraphics[width=1.5in]{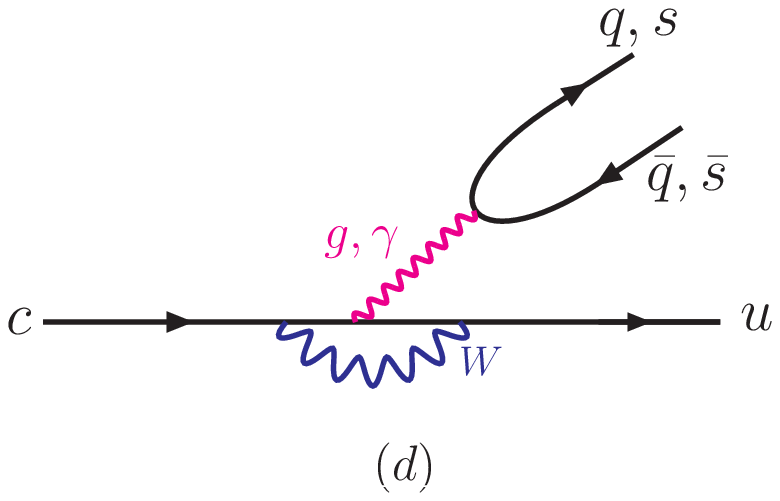}

\caption{Feynman diagrams for the c quark decays, where
(a,b,c) are the tree-level processes 
with the $W$-boson emissions to directly connect to  quark pairs, 
where $q=(d,s)$,
while (d) corresponds to the penguin-level ones
with the $W$-boson in the loop.
}\label{dia}
\end{figure}
 From Fig.~\ref{dia}, 
there are four types of diagrams for the non-leptonic charm quark decays,
where Figs.~\ref{dia}a$-$\ref{dia}c with the $W$-boson emissions 
directly connected to quark pairs are the so-called tree-level processes,
while Fig. \ref{dia}d with the $W$-boson in the loop corresponds to the penguin-level ones.
In Fig.~\ref{dia}c,
the $c\to du\bar s$ transition that 
proceeds through $|V_{cd}V_{us}|\simeq \sin^2\theta_c$
is the doubly Cabibbo-suppressed one, 
with $\theta_c$ being the well-known Cabibbo angle.
Meanwhile, the $c\to u q\bar q(s\bar s)$ transitions in Fig.~\ref{dia}d
have the higher-order contributions from the quark loops, 
with the effective Wilson coefficients~\cite{Li:2012cfa}
calculated to be smaller than the tree-level ones by one order of magnitude.
As a result, the decay processes in Figs.~\ref{dia}c and \ref{dia}d
are both excluded in the present study.
Accordingly, the effective Hamiltonian 
for the $c\to  s u\bar d$ and $c\to u q\bar q$ transitions with $q=(d,s)$
in Figs.~\ref{dia}a and ~\ref{dia}b, respectively, is given by~\cite{Buras:1998raa}
\begin{eqnarray}\label{Heff}
{\cal H}_{eff}&=&\frac{G_F}{\sqrt 2}
\bigg\{V_{cs}V_{ud} [c_1 O_1+c_2 O_2]+\sum_{q=d,s}
V_{cq}V_{uq} [c_1 O_1^q+c_2 O_2^q]\bigg\}\,,
\end{eqnarray}
with the current-current operators $O_{1,2}^{(q)}$, written as
\begin{eqnarray}\label{O12}
&&
O_1=(\bar u d)_{V-A}(\bar s c)_{V-A}\,,\;
O_2=(\bar s d)_{V-A}(\bar u c)_{V-A}\,,\nonumber\\
&&
O_1^q=(\bar u q)_{V-A}(\bar q c)_{V-A}\,,\;
O_2^q=(\bar q q)_{V-A}(\bar u c)_{V-A}\,,
\end{eqnarray}
where $G_F$ is the Fermi constant, $V_{ij}$ are
the Cabibbo-Kobayashi-Maskawa (CKM) matrix elements, and
$(\bar q_1 q_2)_{V-A}$ stands for $\bar q_1\gamma_\mu(1-\gamma_5)q_2$.
The operators $O_{1,2}$ and $O_{1,2}^q$ in Eq.~(\ref{Heff})  
lead to the so-called Cabibbo-allowed and Cabibbo-suppressed 
decay modes due to the factor of 
$|(V_{cq}V_{uq})/(V_{cs}V_{ud})|=\sin\theta_c$.
%
%
The Wilson coefficients $c_{1,2}$ in Eq.~(\ref{Heff}) are scale-dependent.
In the NDR scheme~\cite{Fajfer:2002gp,Li:2012cfa},
one has that $(c_1,c_2)=(1.27,-0.51)$ at the scale $\mu=1$ GeV. 
Note that one is able to recombine
$V_{cs}V_{ud}[c_1 O_1+c_2 O_2]$ and 
$\sum_{q=d,s}V_{cq}V_{uq} [c_1 O_1^q+c_2 O_2^q]$ in Eq.~(\ref{Heff}) into 
$V_{cs}V_{ud}[c_+ O_+ +c_- O_-]$ and 
$V_{cd}V_{ud}[c_+\hat O_+ +c_- \hat O_-]$
with $\hat O_\pm \equiv O_\pm^d-O_\pm^s$, respectively,
where $c_\pm=c_1\pm c_2$, 
$O_\pm^{(q)}=(O_1^{(q)}\pm O_2^{(q)})/2$ and 
$V_{cs}V_{us}=-V_{cd}V_{ud}$.

For the four-quark operator $(\bar q^i q_k)(\bar q^j c)$ from the effective Hamiltonian
 in Eq.~(\ref{O12}), $\bar q^i q_k \bar q^j$ that 
belongs to the $SU(3)$ triplet of $q_i=(u,d,s)$ can be decomposed as the irreducible forms of
$\bar 3\times 3\times \bar 3=\bar 3+\bar 3'+6+\overline{15}$,
which are in terms of the $SU(3)$ flavor symmetry
with the Lorentz-Dirac structures being disregarded. 
Consequently,
the Cabibbo-allowed operators $O_-$ and $O_+$
fall into 6 and $\overline{15}$, respectively, instead of $\bar 3+\bar 3'$
that actually appear in the penguin operators. 
Therefore, in the $SU(3)$ picture
the Cabibbo-allowed operators $O_-$ and $O_+$
are presented as~\cite{Savage:1989qr,Savage:1991wu}
\begin{eqnarray}
{\cal O}_6&=&{1\over 2}[(\bar u d)(\bar s c)-(\bar s d)(\bar u c)]\,,\nonumber\\
{\cal O}_{\overline{15}}&=&{1\over 2}[(\bar u d)(\bar s c)+(\bar s d)(\bar u c)]\,,
%
\end{eqnarray}
which are formed as the tensor notations of 
$H(6)^{ij}$ and $H(\overline{15})^i_{jk}$, 
respectively. Note that the Cabibbo-suppressed operators 
$\hat O_-$ and $\hat O_+$ have similar  irreducible forms, leading to their own 
$\hat H(6)^{ij}$ and $\hat H(\overline{15})^i_{jk}$~\cite{Savage:1989qr,Savage:1991wu}.
%
As a result,
the effective Hamiltonian in Eq.~(\ref{Heff}) under the $SU(3)$ representation 
becomes 
\begin{eqnarray}
{\cal H}_{eff}=\frac{G_F}{\sqrt 2}
\bigg\{V_{cs}V_{ud} [c_- H(6)+c_+ H(\overline{15})]
+V_{cd}V_{ud}[c_-\hat H(6)+c_+\hat H(\overline{15})]\bigg\}\,,
\end{eqnarray}
where the non-zero entries are
\begin{eqnarray}
&&H^{22}(6)=2\,,H^2_{13}(\overline{15})=H^2_{31}(\overline{15})=1\,,\nonumber\\
&&\hat H^{23}(6)=\hat H^{32}(6)=-2\,,\nonumber\\
&&\hat H^2_{12}(\overline{15})=\hat H^2_{21}(\overline{15})=
-\hat H^3_{13}(\overline{15})=-\hat H^3_{31}(\overline{15})=1\,.
\end{eqnarray}
%
%

To proceed, we take the amplitudes of ${\bf B}_c\to {\bf B}_n M$ 
under the $SU(3)$ representations. 
First, the ${\bf B}_c$ state 
acts as $\bar 3$ under the $SU(3)$ flavor symmetry, written as
\begin{eqnarray}
({\bf B}_c)^i&=&(\Xi_c^0,-\Xi_c^+,\Lambda_c^+)\,,
\end{eqnarray}
by which one defines $T_{ij}=\epsilon_{ijk}({\bf B}_c)^k$.
Second,
${\bf B}_n$ is the baryon octet, given by
\begin{eqnarray}
({\bf B}_n)^i_j&=&\left(\begin{array}{ccc} 
\frac{1}{\sqrt{6}}\Lambda+\frac{1}{\sqrt{2}}\Sigma^0 & \Sigma^+ & p\\
 \Sigma^- &\frac{1}{\sqrt{6}}\Lambda -\frac{1}{\sqrt{2}}\Sigma^0  & n\\
 \Xi^- & \Xi^0 &-\sqrt{\frac{2}{3}}\Lambda 
\end{array}\right)\,.
\end{eqnarray}
To include the octet $(\pi,K,\eta_8)$ and
singlet $\eta_1$, $M$ is presented as the nonet, given by
\begin{eqnarray}
(M)^i_j=\left(\begin{array}{ccc} 
\frac{1}{\sqrt{2}}(\pi^0+ \cos\phi\eta +\sin\phi\eta' ) & \pi^- & K^-\\
 \pi^+ & \frac{-1}{\sqrt{2}}(\pi^0- \cos\phi\eta -\sin\phi\eta') & \bar K^0\\
 K^+ & K^0& -\sin\phi\eta +\cos\phi\eta' 
\end{array}\right)\,,
\end{eqnarray}
where $(\eta,\eta')$ are the mixtures of $(\eta_1,\eta_8)$,
decomposed as 
$\eta_1=\sqrt{2/3}\eta_q+\sqrt{1/3}\eta_s$ and
$\eta_8=\sqrt{1/3}\eta_q-\sqrt{2/3}\eta_s$
with $\eta_q=\sqrt{1/2}(u\bar u+d\bar d)$ and $\eta_s=s\bar s$.
Explicitly, the $\eta-\eta'$ mixing matrix is given by~\cite{FKS}
\begin{eqnarray}\label{eta_mixing}
\left(\begin{array}{c} \eta \\ \eta^\prime \end{array}\right)
=
\left(\begin{array}{cc} \cos\phi & -\sin\phi \\ \sin\phi & \cos\phi \end{array}\right)
\left(\begin{array}{c} \eta_q \\ \eta_s \end{array}\right),
\end{eqnarray}
with the mixing angle $\phi=(39.3\pm1.0)^\circ$.
%
\begin{table}[b]
\caption{The tree-level amplitudes for the $\Lambda_c^+\to {\bf B}_n M$ decays.}\label{tab1}
\begin{tabular}{|c||c|c|}
\hline 
Decay modes&$T({\cal O}_{\overline{15}})$&$T({\cal O}_6)$  \\
\hline
$T(\Lambda_c^+ \to p \bar K^0)$ 
&$a+c$ & $-2e$\\
$T(\Lambda_c^+ \to \Lambda \pi^+)$  
&$\sqrt{\frac{1}{6}}(a+b-2c)$
& $-\sqrt{\frac{2}{3}}(e+f+g)$\\
$T(\Lambda_c^+ \to \Sigma^+ \pi^0)$  
&$-\sqrt{\frac{1}{2}}(a-b)$& $\sqrt{2}(e-f-g)$ \\
$T(\Lambda_c^+ \to \Sigma^0 \pi^+)$ 
&$\sqrt{\frac{1}{2}}(a-b)$& $-\sqrt{2}(e-f-g)$\\
$T(\Lambda_c^+ \to \Xi^0 K^+)$ 
&$b+ d$& $-2f$ \\ 
\hline
$\hat T(\Lambda_c^+ \to p \pi^0)$ 
&$-\sqrt{\frac{1}{2}}(b+c)$&  $\sqrt{2}(f+g)$ \\
$\hat T(\Lambda_c^+ \to \Lambda K^+)$  
&$\sqrt{\frac{1}{6}}(-a+2b+2c+3d)$
&$ \sqrt{\frac{2}{3}}(e-2f+g)$\\
$\hat T(\Lambda_c^+ \to \Sigma^0 K^+)$   
&$-\sqrt{\frac{1}{2}}(a+d)$& $\sqrt{2}(e-g)$ \\
$\hat T(\Lambda_c^+ \to \Sigma^+ K^0)$  
&$(-a+d)$& $2(e-g)$ \\
\hline 
$T(\Lambda_c^+ \to \Sigma^+ \eta)$  
&$-d\sin\phi+\sqrt{\frac{1}{2}}(a+b)\cos\phi$ & $-\sqrt{2}[e+(f-g)]\cos\phi$     \\
&$+h_1(\sqrt{2}\cos\phi-\sin\phi)$&$-2h_2(\sqrt{2}\cos\phi-\sin\phi)$\\
$T(\Lambda_c^+ \to \Sigma^+ \eta^\prime)$ 
&$d\cos\phi+\sqrt{\frac{1}{2}}(a+b)\sin\phi$& $-\sqrt{2}[e+(f-g)]\sin\phi$\\
&$+h_1(\cos\phi+\sqrt{2}\sin\phi)$& $-2h_2(\sqrt{2}\sin\phi+\cos\phi)$\\
$\hat T(\Lambda_c^+ \to p \eta)$  
&$-(a+c+d)\sin\phi+\sqrt{\frac{1}{2}}(b-c)\cos\phi$& $-\sqrt{2}[\sqrt{2} e\sin\phi -(f-g)\cos\phi]$\\
&$+h_1(\sqrt{2}\cos\phi-\sin\phi)$&$+2h_2(\sqrt{2}\cos\phi-\sin\phi)$\\
$\hat T(\Lambda_c^+ \to p \eta^\prime)$  
&$(a+c+d)\sin\phi+\sqrt{\frac{1}{2}}(b-c)\cos\phi$&$\sqrt{2}[\sqrt{2}e\cos\phi +(f-g)\sin\phi]$\\
&$+h_1(\cos\phi+\sqrt{2}\sin\phi)$&$+2h_2(\sqrt{2}\sin\phi+\cos\phi)$\\
\hline
\end{tabular}
\end{table}
Subsequently, the amplitude of ${\bf B}_c\to {\bf B}_n M$ is derived as
\begin{eqnarray}\label{amp_SU3}
&&{\cal A}({\bf B}_c\to {\bf B}_n M)=\langle {\bf B}_n M|H_{eff}|{\bf B}_c\rangle\nonumber\\
&=&\frac{G_F}{\sqrt 2}[V_{cs} V_{ud}T({\bf B}_c\to {\bf B}_n M)+
V_{cd}V_{ud}\hat T({\bf B}_c\to {\bf B}_n M)]\,,
\end{eqnarray}
where $T({\bf B}_c\to {\bf B}_n M)=T({\cal O}_{\overline{15}})+T({\cal O}_6)$ 
are given by~\cite{Lu:2016ogy}
\begin{eqnarray}\label{Tq}
T({\cal O}_{\overline{15}})&=&
aH^{i}_{jk}(\overline{15})({\bf B}_{c})^j({\bf B}_n)^k_l (M)^l_i  
+bH^{i}_{jk}(\overline{15})({\bf B}_{c})^j (M)^k_l ({\bf B}_n)^l_i  \nonumber\\
&+&
cH^{i}_{jk}(\overline{15}) ({\bf B}_n)^j_l (M)^k_i ({\bf B}_{c})^l
+dH^{i}_{jk}(\overline{15}) (M)^j_l ({\bf B}_n)^k_i ({\bf B}_{c})^l\nonumber\\
&+&
h_1H^{i}_{jk}(\overline{15})({\bf B}_n)^k_i (M)^l_l ({\bf B}_{c})^j\,,\nonumber\\
T({\cal O}_6)&=&
eH^{ij}(6)T_{ik}({\bf B}_n)^k_l (M)^l_j+
fH^{ij}(6)T_{ik}(M)^k_l ({\bf B}_n)^l_j\nonumber\\
&+&
gH^{ij}(6)({\bf B}_n)^k_i (M)^l_j T_{kl}+
h_2 H^{ij}(6)T_{ik}({\bf B}_n)^k_j (M)^l_l\,,
\end{eqnarray}
with $(c_+,c_-)$ absorbed in the $SU(3)$ parameters of 
$(a,b,c,d,h_1)$ and $(e,f,g,h_2)$, respectively,
while $\hat T({\bf B}_c\to {\bf B}_n M)$ is given by
replacing $H(6,\overline{15})$ in $T({\bf B}_c\to {\bf B}_n M)$ with 
$\hat H(6,\overline{15})$, respectively.
%
Since the amplitudes are derived from the effective Hamiltonian in Eq.~(\ref{Heff}), where
the $c\to  s u\bar d$ and $c\to u q\bar q$ transitions are the tree-level processes,
$T({\bf B}_c\to {\bf B}_n M)$ and $\hat T({\bf B}_c\to {\bf B}_n M)$
are named as the tree-level ($T$) amplitudes.
In Eq.~(\ref{Tq}), the expansions of 
$T({\bf B}_c\to {\bf B}_n M)$ and $\hat T({\bf B}_c\to {\bf B}_n M)$ 
are shown in Table~\ref{tab1}.
 Note that, although we follow the approach in Ref.~\cite{Savage:1991wu},
the $h_{1,2}$ terms are newly added for the singlet $\eta_1$.
Due to $c_-/c_+\simeq 2.4$, the contribution of ${\cal O}_6(\hat {\cal O}_6)$ to 
the decay branching ratio can be 5.5 times larger than that of 
${\cal O}_{\overline{15}}(\hat {\cal O}_{\overline{15}})$, 
such that ${\cal O}_{\overline{15}}(\hat {\cal O}_{\overline{15}})$ is negligible.
However, we will examine if the reduction is reasonable, in case
the interferences between ${\cal O}_6(\hat {\cal O}_6)$ and 
${\cal O}_{\overline{15}}(\hat {\cal O}_{\overline{15}})$ can be sizable.
Subsequently, we only keep the $SU(3)$ parameters $e$, $f$, $g$ and $h_2$ 
from ${\cal O}_6$ to simplify the amplitudes.
Since $e$, $f$, $g$ and $h_2$ 
are complex numbers, we have 7 real independent parameters
to be determined by the data, given by
\begin{eqnarray}\label{7p}
e, fe^{i\delta_f},ge^{i\delta_g},h_2e^{i\delta_{h_2}}\,,
\end{eqnarray}
where $e$ is set to be real, while 
an overall phase can be removed without losing generality.
To calculate the decay widths,
we use~\cite{pdg}:
\begin{eqnarray}
\Gamma(\Lambda_c\to {\bf B}_n M)=
\frac{|\vec{p}_{cm}|}{8\pi m_{\Lambda_c}^2}|{\cal A}(\Lambda_c\to {\bf B}_n M)|^2\,,
\end{eqnarray}
where $|\vec{p}_{cm}|=
\sqrt{[(m_{\Lambda_c}^2-(m_{{\bf B}_n}+m_M)^2]
[(m_{\Lambda_c}^2-(m_{{\bf B}_n}-m_M)^2]}/(2 m_{\Lambda_c})$, 
with the integrated-over variables of the phase spaces in the two-body decays.

\section{Numerical Results and Discussions }
For the numerical analysis,
we use the minimum $\chi^2$ fit to find 
the $SU(3)$ parameters in Eq.~(\ref{7p}). 
The theoretical inputs for the CKM matrix elements are given by~\cite{pdg}
\begin{eqnarray}\label{B1}
&&(V_{cs},V_{ud},V_{us},V_{cd})=(1-\lambda^2/2,1-\lambda^2/2,\lambda,-\lambda)\,,
\end{eqnarray}
with $\lambda=0.225$ in the Wolfenstein parameterization.
There are 9 branching ratios of $\Lambda_c\to B_n M$,
which are the data inputs, given in the last column of Table~\ref{tab2}. 
The equation of the $\chi^2$ fit is given by
\begin{eqnarray}
\chi^2=\sum_{i=1}^{9} \bigg(\frac{{\cal B}^i_{th}-{\cal B}^i_{ex}}{\sigma_{ex}^i}\bigg)^2\,,
\end{eqnarray}
where ${\cal B}_{th}^i$ and ${\cal B}_{ex}^i$ stand for
the branching ratios from the theoretical $SU(3)$ amplitudes in Table~\ref{tab1}
and experimental data inputs in Table~\ref{tab2},  
with $\sigma^i_{ex}$ as the 1$\sigma$ experimental errors, while
$i=1, 2, ..., 9$ denote the 9 observed decay modes involved in the global fit, respectively.
Consequently, we obtain
\begin{eqnarray}\label{efgh}
(e,f,g,h_2)&=&
(0.257\pm 0.006,0.121\pm 0.015,0.092\pm 0.021,0.111\pm 0.081)\,\text{GeV}^3\,,\nonumber\\
(\delta_f,\delta_g,\delta_{h_2})&=&(79.0\pm 6.8, 35.2\pm 8.8,102.4\pm 29.8)^\circ\,,\nonumber\\
\chi^2/d.o.f&=&2.4\,,
\end{eqnarray}
where $d.o.f$ stands for the degrees of freedom.
The statistical p-value to be smaller than 0.05 will show 
the inconsistency between the theory and data~\cite{pdg}, 
which is equivalent  to $\chi^2/d.o.f>3$ here. In our case, 
the value of $\chi^2/d.o.f=2.4$ indicates a tolerable result 
to accommodate the current data of ${\cal B}(\Lambda_c\to {\bf B}_n M)$
under the $SU(3)$ flavor symmetry, 
 where the contributions of ${\cal O}_{\overline{15}}(\hat {\cal O}_{\overline{15}})$
and the broken effects of $SU(3)$ are both neglected.
Explicitly, it is found that the largest contributions to $\chi^2$ are from 
${\cal B}(\Lambda_c^+\to \Lambda K^+,\Sigma^0 K^+)$,
whereas the individual $\chi^2$ values from the other seven data 
show no apparent violation of SU(3) 
or the sextet dominating assumption.
The value of $\chi^2/d.o.f.$ being as large as 2.4
could suggest that the decays of $\Lambda_c^+\to (\Lambda K^+,\Sigma^0 K^+)$
should be reexamined by BESIII with  more precisions.
Due to the lack of sufficient data,
it leaves the room for  more precise examinations by the future experimental measurements
on the $SU(3)$ flavor symmetry 
with or without
${\cal O}_{\overline{15}}(\hat {\cal O}_{\overline{15}})$.
%
With the parameters in Eq.~(\ref{efgh}) we can reproduce 
the branching ratios of the measured two-body $\Lambda_c$ decays 
as shown in Table~\ref{tab2}, where the results based on
the heavy quark effective theory (HQET)~\cite{Sharma:1998rd}, 
Sharma and Verma (SV) in Ref.~\cite{Sharma:1996sc}, 
pole model (PM)~\cite{Cheng:1993gf} and 
current 
algebra (CA)~\cite{Cheng:1993gf} are also listed.
In our fit,
the $SU(3)$ amplitudes in Eq.~(\ref{efgh}) have 
considerable imaginary parts, being included in
$\delta_{f,g,h_2}$. Nonetheless, the studies 
in Refs.~\cite{Zenczykowski:1993hw,Sharma:1998rd} 
depend on real ones.
%
%
For a test, we turn off $\delta_{f,g,h_2}$, 
which causes an unsatisfactory fit to the data with 
$\chi^2/d.o.f\approx 14\gg 2.4$ in Eq.~(\ref{efgh}),
suggesting that the imaginary parts are necessary
to fit the nine data well.
It is similar that, in the $D\to MM$ decays, 
the imaginary parts with the $SU(3)$ flavor symmetry
are also considerable, 
which correspond to the strong phases calculated from
the on-shell quark loops in the next-leading-order 
QCD models~\cite{Cheng:2012xb}.
We hence conclude that the $\Lambda_c^+\to {\bf B}_n M$ decays
are like the $D\to MM$ ones, where the phases are 
in accordance with the higher order contributions 
in the QCD models, which have not been well developed yet.

\begin{table}[b]
\caption{The branching ratios of the $\Lambda_b\to {\bf B}_n M$ decays,
where the 2nd column  is for our results,  
where the errors come from the parameters in Eq.~(\ref{efgh}),
while  3, 4, ..., 7 ones correspond to  the studies by
the heavy quark effective theory (HQET)~\cite{Sharma:1998rd}, 
Sharma and Verma (SV) in Ref.~\cite{Sharma:1996sc}, 
pole model (PM)~\cite{Cheng:1993gf},
current algbra (CA)~\cite{Cheng:1993gf} 
and  data~\cite{pdg,Ablikim:2015flg,Ablikim:2017ors}, respectively.}\label{tab2}
\begin{tabular}{|c|ccccc|c|}
\hline 
Branching ratios&Our results
&HQET~\cite{Sharma:1998rd}
&SV~\cite{Sharma:1996sc}
&PM~\cite{Cheng:1993gf}
&CA~\cite{Cheng:1993gf}
&Data~\cite{pdg,Ablikim:2015flg,Ablikim:2017ors} \\
\hline
$10^2{\cal B}(\Lambda_c^+ \to p \bar K^0)$ 
& $3.3\pm 0.2$
&$1.23$
&$2.67\pm 0.74$
&$1.20$
&$3.46$
&$3.16\pm 0.16$\\ 
$10^2{\cal B}(\Lambda_c^+ \to \Lambda \pi^+)$  
& $1.3\pm 0.2$
& $1.17$
&-----
&$0.84$
&$1.39$
&$1.30\pm 0.07$\\
$10^2{\cal B}(\Lambda_c^+ \to \Sigma^+ \pi^0)$  
& $1.3\pm 0.2$
& $0.69$
&-----
&$0.68$
&$1.67$
& $1.24\pm 0.10$\\
$10^2{\cal B}(\Lambda_c^+ \to \Sigma^0 \pi^+)$ 
& $1.3\pm 0.2$ 
&$0.69$
&$0.87\pm 0.20$
&$0.68$
&$1.67$
& $1.29\pm 0.07$\\
$10^2{\cal B}(\Lambda_c^+ \to \Xi^0 K^+)$ 
& $0.5\pm 0.1$ 
& $0.07$
&-----
&-----
&-----
& $0.50\pm 0.12$\\
\hline\hline
$10^4{\cal B}(\Lambda_c^+ \to p \pi^0)$ 
&  $5.6\pm 1.5$
&-----
&$2$
&-----
&-----
&----- \\
$10^4{\cal B}(\Lambda_c^+ \to \Lambda K^+)$  
&$4.6\pm 0.9$
&-----
&14
&-----
&-----
&$6.1\pm 1.2$\\
$10^4{\cal B}(\Lambda_c^+ \to \Sigma^0 K^+)$   
& $4.0\pm 0.8$
&-----
&4
&-----
&-----
&$5.2\pm 0.8$ \\
$10^4{\cal B}(\Lambda_c^+ \to \Sigma^+ K^0)$  
& $8.0\pm 1.6$
&-----
&9
&-----
&-----
&----- \\
\hline\hline
$10^2{\cal B}(\Lambda_c^+ \to \Sigma^+ \eta)$  
& $0.7\pm 0.4$
& $0.25$
& $0.50\pm 0.17$
&-----
&-----
&$0.70\pm 0.23$ \\
$10^2{\cal B}(\Lambda_c^+ \to \Sigma^+ \eta^\prime)$ 
& $1.0^ {+1.6}_{-0.8}$
& $0.08$
& $0.20\pm 0.08$
&-----
&-----
&-----  \\
$10^4{\cal B}(\Lambda_c^+ \to p \eta)$  
& $12.4\pm 4.1$
&-----
&21
&-----
&-----
&$12.4\pm 3.0$ \\
$10^4{\cal B}(\Lambda_c^+ \to p \eta^\prime)$  
& $12.2^{+14.3}_{-\,\,\,8.7}$
&-----
&4
&-----
&-----
&-----  \\
\hline
\end{tabular}
\end{table}

From Table~\ref{tab1}, 
 by keeping both ${\cal O}_{6}$ and ${\cal O}_{\overline{15}}$,
we obtain that
\begin{eqnarray}\label{re_A0}
{\cal A}(\Lambda_c^+\to \Sigma^0 \pi^+)&=&-{\cal A}(\Lambda_c^+\to \Sigma^+ \pi^0)\,,\nonumber\\
{\cal A}(\Lambda_c^+\to \Sigma^+ K^0)&=&\sqrt 2 {\cal A}(\Lambda_c^+\to \Sigma^0 K^+)\,,
\end{eqnarray}
where $\Lambda_c^+\to \Sigma^0 \pi^+$ being identical to 
$\Lambda_c^+\to\Sigma^+ \pi^0$ represents the conservation of 
the isospin ($SU(2)$) symmetry.
By neglecting ${\cal O}_{\overline{15}}$, the relations in Eq.~(\ref{re_A0}) can be extended to
\begin{eqnarray}\label{re_A}
{\cal A}(\Lambda_c^+\to \Sigma^0 \pi^+)&=&-{\cal A}(\Lambda_c^+\to \Sigma^+ \pi^0)\,,\nonumber\\
{\cal A}(\Lambda_c^+\to \Sigma^+ K^0)&=&\sqrt 2 {\cal A}(\Lambda_c^+\to \Sigma^0 K^+)\,,\nonumber\\
\sqrt 6 {\cal A}(\Lambda_c^+\to \Lambda \pi^+)&+&\sqrt 2 {\cal A}(\Lambda_c^+\to \Sigma^0 \pi^+)
=2{\cal A}(\Lambda_c^+\to p \bar K^0)\,,\nonumber\\
%
%
\sqrt 6 {\cal A}(\Lambda_c^+\to \Lambda \pi^+)&-&\sqrt 2 {\cal A}(\Lambda_c^+\to \Sigma^0 \pi^+)
=\frac{2\sqrt 2}{\lambda}{\cal A}(\Lambda_c^+\to p \pi^0)\,,
\end{eqnarray}
resulting in
\begin{eqnarray}\label{re_B}
{\cal B}(\Lambda_c^+\to \Sigma^0 \pi^+)&=& {\cal B}(\Lambda_c^+\to \Sigma^+ \pi^0)\,,\nonumber\\
{\cal B}(\Lambda_c^+\to \Sigma^+ K^0)&=&2{\cal B}(\Lambda_c^+\to \Sigma^0 K^+) \,,\nonumber\\
{\cal B}(\Lambda_c^+\to p \pi^0)&\simeq &\frac{\lambda^2}{2}
[3{\cal B}(\Lambda_c^+\to \Lambda \pi^+)+{\cal B}(\Lambda_c^+\to \Sigma^0 \pi^+)-
{\cal B}(\Lambda_c^+\to p \bar K^0)]\,.
\end{eqnarray}
With the inputs of the $SU(3)$ parameters in Eq.~(\ref{efgh}),
we show  
${\cal B}(\Lambda_c^+ \to \Sigma^+ K^0)=(8.0\pm 1.6)\times 10^{-4}$ and
${\cal B}(\Lambda_c^+\to \Sigma^0 K^+)=(4.0\pm 0.8)\times 10^{-4}$
to agree with the second relation in Eq.~(\ref{re_B}), which can be used
to test the assumption of the dominant ${\cal O}_6$ contributions
in comparison with the future measurements.
We remark that the factorization approach predicts
${\cal B}(\Lambda_c^+\to \Sigma^0 \pi^+,\Sigma^0 K^+)\simeq 0$,
which contradicts the relations from the $SU(3)$ symmetry.
This is due to the fact that, 
when the decay proceeds with the $\Lambda_c^+\to \Sigma^0$ transition,
together with the recoiled meson $\pi^+$ or $K^+$, 
the $c\to s$ transition currents transform
$\Lambda_c^+\to \Lambda=(ud-du)s$, which is unable to correlate to 
$\Sigma^0=(ud+du)s$~\cite{Hsiao:2017tif}, 
leading to ${\cal B}=0$.

In Eq.~(\ref{re_B}),
the simple estimation based on the data inputs
gives that ${\cal B}(\Lambda_c^+\to p \pi^0)=(5.1\pm 0.7)\times 10^{-4}$, 
which agrees with our numerical fitting result of  
${\cal B}(\Lambda_c^+\to p\pi^0)=(5.6\pm 1.5)\times 10^{-4}$, but is larger
than the experimental upper bound of $3\times 10^{-4}$ (90\%C.L.) in Eq.~(\ref{data1b}) by BESIII.
To check if there is a discrepancy
here, we have taken the original data
of ${\cal B}(\Lambda_c^+\to p\pi^0)=(7.95\pm 13.61)\times 10^{-5}$~\cite{ppi0} by BESIII
as the  input. In this case, we get $\chi^2/d.o.f=4.7$, which is
two times larger than the value in Eq.~(\ref{efgh}),
showing that the fitting
cannot accommodate the present data of ${\cal B}(\Lambda_c^+\to p\pi^0)$.
Apart from the $SU(3)$ flavor symmetry,
we estimate that ${\cal B}(\Lambda_c^+ \to p \pi^0)\simeq 5\times 10^{-4}$
in the approach of the factorization, 
which is also larger than the experimental upper bound.
It is clear that a dedicated search for this mode with a more precise measurement should
be done.
An improved sensitivity to measure $\Lambda_c^+\to p\pi^0$
will clarify if the currently unmovable discrepancy
exists or not.

It is also interesting to see that 
${\cal B}(\Lambda_c^+ \to \Sigma^+ \eta^\prime)$ and 
${\cal B}(\Lambda_c^+ \to p \eta^\prime)$ fitted to be 
$(1.0^{+1.6}_{-0.8})\times 10^{-2}$ and $(12.2^{+14.3}_{-\,\,\,8.7})\times 10^{-4}$
are as large as their $\eta$ counterparts,  respectively, while 
${\cal B}(\Lambda_b\to \Lambda\eta)\simeq {\cal B}(\Lambda_b\to\Lambda\eta')$~\cite{Geng:2016gul}.
We note that there is a similar term in Ref.~\cite{h_term} 
as the $h_{1,2}$ terms,
which relates $\Lambda_c^+\to \Sigma^+\eta'$ to $\Xi_c^0\to \Xi^0\eta'$.
In contrast, the theoretical approach in Ref.~\cite{Sharma:1996sc} is
based on the $SU(3)$ flavor symmetry also,
but without the 
$h_{1,2}$ terms
to include the singlet $\eta_1$, such that
it leads to ${\cal B}(\Lambda_c^+ \to \Sigma^+(p) \eta^\prime)
<{\cal B}(\Lambda_c^+ \to \Sigma^+(p) \eta)$~\cite{Sharma:1996sc}.
Finally, we remark that, with the  $SU(3)$ symmetry,
we can extend our study to the two-body $\Xi_b^{+,0}$ decays,
which are also accessible to the current experiments.
Since the two-body $\Lambda_c^+\to {\bf B}_nV$ 
with $V$  the vector meson and
three-body $\Lambda_c^+$ decays are observed,
which require the interpretations, 
the approach of the $SU(3)$ symmetry can be useful.

\section{Conclusions}
We have studied the two-body $\Lambda_c^+\to {\bf B}_n M$ decays,
which have been recently reanalyzed or newly measured by BESIII.
With the $SU(3)$ flavor symmetry, we can 
describe
the data except that for $\Lambda_c^+\to p\pi^0$.
We have found that ${\cal B}(\Lambda_c^+\to p\pi^0)=(5.6\pm 1.5)\times 10^{-4}$, 
which is almost $2\sigma$ above the experimental upper bound of
$3\times10^{-4}$.
We hope that the future experimental measurement of ${\cal B}(\Lambda_c^+\to p\pi^0)$
can resolve  this discrepancy.
Unlike the previous results, we have predicted that
${\cal B}(\Lambda_c^+ \to \Sigma^+ \eta^\prime)=(1.0^{+1.6}_{-0.8})\times 10^{-2}$ and 
${\cal B}(\Lambda_c^+ \to p \eta^\prime)=(12.2^{+14.3}_{-\,\,\,8.7})\times 10^{-4}$
which are as large as their $\eta$ counterparts,
due to the newly added $h_{1,2}$ terms with the singlet $\eta_1$
in the $SU(3)$ flavor symmetry.
With the  $SU(3)$ symmetry,
one is able to study $\Lambda_c^+\to {\bf B}_nV$ and
the three-body ${\bf B}_c$ decays, which have been observed but barely interpreted.
Moreover, the extensions to study the $\Xi_b^{+,0}$ decays are possible,
which are also accessible to the current experiments.

\section*{ACKNOWLEDGMENTS}
We would like to thank X.G. He for useful discussions.
This work was supported in part by National Center for Theoretical Sciences,
MoST (MoST-104-2112-M-007-003-MY3), and
National Science Foundation of China (11675030).

\end{document}